\documentclass[english,aps,manuscript]{revtex4}
\usepackage[T1]{fontenc}
\usepackage[latin1]{inputenc}
\usepackage{babel}
\usepackage{graphics}

\makeatletter

\providecommand{\LyX}{L\kern-.1667em\lower.25em\hbox{Y}\kern-.125emX\@}
\let\SF@@footnote\footnote
\def\footnote{\ifx\protect\@typeset@protect
    \expandafter\SF@@footnote
  \else
    \expandafter\SF@gobble@opt
  \fi
}
\expandafter\def\csname SF@gobble@opt \endcsname{\@ifnextchar[%]
  \SF@gobble@twobracket
  \@gobble
}
\edef\SF@gobble@opt{\noexpand\protect
  \expandafter\noexpand\csname SF@gobble@opt \endcsname}
\def\SF@gobble@twobracket[#1]#2{}

\usepackage[T1]{fontenc}

\makeatletter
\def\be{\begin{equation}}
\def\ee{\end{equation}}
\def\bea{\begin{eqnarray}}
\def\eea{\end{eqnarray}}

\makeatother
\begin{document}

\title{Robustness of the Quintessence Scenario in Particle Cosmologies }

\author{Greg Huey and Reza Tavakol}

\affiliation{Astronomy Unit\\
School of Mathematical Sciences\\
Queen Mary, University of London\\
Mile End Road, London E1 4NS, England}
\vskip 1.5cm

\begin{abstract}

We study the robustness of the 
quintessence tracking scenario in the
context of more general
cosmological models that derive from high-energy physics.
We consider the effects of inclusion of
multiple scalar fields, corrections to the Hubble expansion law (such as those
that arise in brane cosmological models), and potentials that decay
with expansion of the Universe. We find that in a successful tracking quintessence model
the average equation of state must remain nearly constant. 
Overall, the conditions for successful tracking
become more complex in these more general settings. 
Tracking can become more fragile
in presence of multiple scalar fields, and more stable
when temperature dependent potentials are present.
Interestingly though, most of the cases where tracking is disrupted
are those in which the cosmological model is itself non-viable due to other
constraints. In this sense tracking remains
robust in models that are cosmologically viable.

\end{abstract}
\maketitle

\section{Introduction}

Recent measurements of the cosmic microwave background (CMB) anisotropy
power spectrum by the MAT~\cite{QMAP-MAT-TOCO}, BOOMERANG~\cite{BOOMERANG}
and MAXIMA~\cite{MAXIMA} experiments suggest the Universe is nearly
spatially flat, with a total density, \( \Omega _{tot0} \),
very close to unity. At the same time, there is ample evidence
from observations, including CMB power spectrum~\cite{QMAP-MAT-TOCO},
galaxy clustering statistics~\cite{GlxyClstrStat}, peculiar velocities~\cite{PecVeloc}
and the baryon mass fraction in clusters of galaxies~\cite{BrynFrac,StrFormBound}
that the density of the clumped (ie: baryonic and dark) matter in
the Universe is substantially lower, being of the order of 30\% of
the critical value (\( \Omega _{matter0}\sim 0.3 \)). Additionally the
evidence from the spectral and photometric observations of Type Ia
Supernovae~\cite{perl} seem to suggest that the Universe is undergoing
accelerated expansion at the present epoch (see e.g.~\cite{perl}
and references therein).

One way of explaining this seemingly diverse set of observations is
to postulate that a substantial proportion of the energy density of
the Universe is in the form of a dark component that makes up the
difference between the critical and matter energy densities, which is smooth
on cosmological scales and which possesses a negative pressure. Various
alternatives have been put forward as candidates for this dark component.
One such candidate is a cosmological constant. This choice, however,
involves an undesirable fine tuning problem, in that the ratio
of the cosmological constant and the matter energy densities in the
early universe need to be set to an infinitesimal value to ensure
their near-coincidence at the present epoch.

An alternative - and arguably more attractive - 
candidate is quintessence.
Though little is known about the actual composition of quintessence,
it has been shown that, should it exist, quintessence can in general
be modeled as a scalar field rolling in a potential~\cite{PJS_Ori_quint},
an approach we adopt here.
Additionally,
it has been shown that some scalar field quintessence models have
the appealing property of possessing attractor-type solutions for which the quintessence
energy density closely tracks the energy density of the rest of the
universe through most of its history~\cite{RPtrack,PJStrack}. The presence of such
attractor-type solutions implies that their asymptotic behavior is largely independent
of initial conditions. This allows the quintessence energy in the early
universe to be comparable to that of
the rest of the universe, thus providing the possibility of 
removing the fine-tuning problem that exists with
the cosmological constant.

Ultimately, any successful cosmological model must be based on a theory
of high energy physics, such as the string theory, M-theory or supergravity.
These models are much more complex than is generally allowed
for in usual studies of tracking. For tracking quintessence to
be truly free from fine-tuning problems, the tracking phenomena must
be robust in the more complicated and complete cosmological settings
that derive from such high energy physics theories. Such scenarios
typically include many scalar fields (some of which may not have potentials
suitable for tracking), a Hubble law different from that of the 
Friedmann-Lemaitre-Robertson-Walker (FLRW)
model at high energies and scalar field effective potentials that
depend explicitly on the scale factor (or equivalently, temperature).
Tracking would be of limited use as a solution to the fine-tuning
problem if it was destroyed by the inclusion of such effects. 

The aim of this
paper is to study the robustness of tracking in a more complete framework,
by considering the effects of the three types of generalizations mentioned
above. Though these additions make the resulting cosmological model more complex,
it is important to consider their effects because in a realistic cosmology
motivated by high-energy physics they are likely to be present, and
because their inclusion leads to qualitatively new effects as well as new
constraints.

The structure of the paper is as follows. In section \ref{EQNS-FIXED} we begin with
the equations of motion for \( n \)~scalar fields, with possible
scale-factor dependent potentials as well as a non-FLRW expansion
rate, and derive the corresponding fixed points. In section \ref{sectioniii} we discuss how the
tracking attractor-type solutions can be understood as a shadowing of instantaneous
fixed points. Section \ref{sectioniv} contains our analysis of  the stability of this shadowing and the independence
of the attractors from the initial conditions. Finally
section \ref{sectionv} contains our conclusions.
%----------------------------------------------
\section{Equations of motion and fixed points}
\label{EQNS-FIXED}
%----------------------------------------------
In a general cosmological scenario inspired by high-energy physics,
one expects the presence of a number of additional
ingredients, among them: multiple scalar fields,
generalized Hubble expansion laws and potentials that decay with expansion.
Multiple scalar fields (\( \phi _{i} \), \( i=1,\ldots ,n \)) arise 
naturally in theories of high energy physics.
Generalized Hubble expansion laws (with \( H_{gen} \))
different from the standard  FLRW Hubble law \( H_{FLRW} \) 
can arise from corrections arising
from the specific model one is considering. Two concrete examples
being the modified Hubble law appearing in a 
braneworld scenario~\cite{BrnWrlModHL,HueyLidsey_BrnInfl,mizunoetal01},
and the effect of varying the strength of gravity.
Explicit dependence of the potential on the scale factor 
can come
about as an effective potential due to the interaction of \( \phi_i  \)
with another field, which has been `integrated out', but
with an energy density that decays with expansion - thus making the effective
potential it induces for \( \phi_i  \) also decay with expansion. 
Here we shall employ potentials of the form
$V_{i}(\phi _{i})\equiv func(\phi _{i})\exp \left( -k_{i}N\right)$, where
$k_i$ are constants and  \( N \) 
is the logarithm of the scale 
factor.
Note that \( k_{i}>0 \) represents
an explicit scale-factor, or equivalently, temperature, dependence
in the potential \( V_{i} \) of the field. This does not allow for
direct coupling between tracking fields, but rather coupling
between tracking fields and the field that has been implicitly integrated out.
The choice of \( k_{i}=0 \) corresponds to the usual form of a
scale factor-independent tracking potential.

The evolution of such cosmological models is governed by the 
equations\footnote{One might question the validity of 
$\ddot{\phi }+3H\dot{\phi }+V_{i,\phi _{i}}=0 $
in such a system. However, note that the 'integrating out' of the
second field $C_{i}$ is done by deriving the full equations of
motion for $\phi _{i}$ and $C_{i}$, solving them and substituting
$ C_{i}(t)$ into the equation of motion for $\phi _{i}$,
then defining a $ V_{i,\phi _{i}}$ such that the $ \phi _{i}$ 
equation of motion has the form $ \ddot{\phi }+3H\dot{\phi }+V_{i,\phi _{i}}=0 $.
Thus, it holds by definition.}

\begin{equation}
\label{Defn_FulSystem}
\begin{array}{c}
\ddot{\phi }_{i}+3H_{gen}\dot{\phi }_{i}+V_{i,\phi _{i}}=0\\
V_{i}(\phi _{i})\equiv func(\phi _{i})\exp \left( -k_{i}N\right) \\
H_{FLRW}\equiv \frac{dN}{dt}=\sqrt{\frac{\kappa ^{2}}{3}\left( \rho _{BG}+\sum _{j}\rho _{j}\right) }\\
\frac{d\ln \left( \rho _{BG}\right) }{dN}=-3\left( 1+w_{BG}\right) =-3\left( \gamma _{BG}\right) 
\end{array}
\end{equation}
 Where \( \rho _{BG} \) is the energy density of the background
which has a constant equation of state \( \gamma _{BG}=1+w_{BG} \),
\( \rho _{j} \) is the energy density of the field \( \phi _{j} \) and 
a dot denotes differentiation with respect to physical time.

To study the possibility of tracking in such systems,
we start by assuming \( H_{gen}=H_{FLRW} \) and 
in order to 
make contact with previous work~\cite{LiddleTrack}, we introduce
the following change of variables
\begin{equation}
\label{Defn_x_y}
\begin{array}{c}
x_{i}\equiv sign(\dot{\phi }_{i})\frac{\sqrt{KE_{i}}}{\sqrt{\rho _{tot}}}=\frac{\dot{\phi }_{i}}{\sqrt{2\rho _{tot}}}\\
y_{i}\equiv \frac{\sqrt{PE_{i}}}{\sqrt{\rho _{tot}}}=\frac{\sqrt{V_{i}(\phi _{i})}}{\sqrt{\rho _{tot}}}.
\end{array}
\end{equation}
The evolution equations then become 
\begin{equation}
\label{xyRhoBg_eom_a}
\begin{array}{c}
\frac{dx_{i}}{dN}=-3x_{i}+\sqrt{\frac{3}{2}}\lambda _{i}y^{2}_{i}+\frac{3}{2}x_{i}\left[ \sum _{j}\left( \left( 1-\frac{k_{j}}{6}\right) 2x^{2}_{j}+\frac{k_{j}}{3}\left( x_{j}^{2}+y^{2}_{j}\right) \right) +\gamma _{BG}\left( 1-\sum _{j}\left( x_{j}^{2}+y^{2}_{j}\right) \right) \right] \\
\frac{dy_{i}}{dN}=-\frac{k_{i}}{2}y_{i}-\sqrt{\frac{3}{2}}\lambda _{i}x_{i}y_{i}+\frac{3}{2}y_{i}\left[ \sum _{j}\left( \left( 1-\frac{k_{j}}{6}\right) 2x^{2}_{j}+\frac{k_{j}}{3}\left( x_{j}^{2}+y^{2}_{j}\right) \right) +\gamma _{BG}\left( 1-\sum _{j}\left( x_{j}^{2}+y^{2}_{j}\right) \right) \right] \\
\frac{d\ln \left( \rho _{BG}\right) }{dN}=-3\left( \gamma _{BG}\right) 
\end{array}
\end{equation}
where \( \lambda _{i}\equiv -\frac{1}{\kappa }\frac{d\ln V_{i}}{d\phi _{i}} \) and 
\( \kappa ^{2}=8\pi G_{N} \).

The generalization to the case of \( H_{gen}\neq H_{FLRW} \)
can be made in the following way.
The instantaneous effect of \( H_{gen}\)
on the evolution of the (\( x_{i},y_{i} \)) variables can be included
through a rescaling which involves replacing \( \lambda _{i} \) in Eq.~(\ref{xyRhoBg_eom_a}) with
\begin{equation}
\label{Defn_lambda_eff}
\widetilde{\lambda }_{i}\equiv \lambda _{i}\frac{H_{FLRW}}{H_{gen}}.
\end{equation}
However, this transformation does not encompass the effect of the
modified expansion rate on the evolution of \( \lambda _{i} \). Thus
such an effect can qualitatively alter the nature of the attractor
solution (see for example~\cite{HueyLidsey_BrnInfl}).
We shall return to this case in  section~\ref{sectioniii}.

In the following we shall also use the physically transparent set of
variables \( \Omega _{\phi i} \) and \( \widehat{\gamma }_{\phi i} \), in terms of which the
above evolution equations become  
\begin{equation}
\label{eom_omegagamma}
\begin{array}{c}
\frac{1}{\Omega _{\phi i}}\frac{d\Omega _{\phi i}}{dN}=
-3\widehat{\gamma }_{\phi i}+
3\left( \gamma _{BG}\left( 1-\sum _{j}\Omega _{\phi j}\right) +\sum _{j}\Omega _{\phi j}\widehat{\gamma }_{\phi j}\right) 
=-3\left( \widehat{\gamma }_{\phi i}-\overline{\gamma }\right) \\
\frac{d\widehat{\gamma }_{\phi i}}{dN}=3\left( \widehat{\gamma }_{\phi i}-\frac{k_{i}}{3}\right) \left( 2-\widehat{\gamma }_{\phi i}\right) \left[ r_{i}-1\right] \; ;\; \; \; r_{i}\equiv \left| \lambda _{i}\right| \sqrt{\frac{\Omega _{\phi i}}{3\left( \widehat{\gamma }_{\phi i}-\frac{k_{i}}{3}\right) \left( 1-\frac{k_{i}}{6}\right) }}
\end{array}
\end{equation}
 where \( \widehat{\gamma }_{\phi i} \) is defined as
\begin{equation}
\label{Defn_gamma_eff}
\widehat{\gamma }_{\phi i}\equiv \left( 1-\frac{k_{i}}{6}\right) \gamma _{\phi i}+\frac{k_{i}}{3}
\end{equation}
and the equation of state for the field \( \phi _{i} \) is defined as \begin{equation}
\label{Defn_eosgamma}
\gamma _{\phi i}\equiv \frac{2x^{2}_{i}}{x^{2}_{i}+y^{2}_{i}}.
\end{equation}
 We also find it useful to define the weighted average of the equation
of state $\overline{\gamma }$, which is the rate of decay of the total energy density of
the universe, in the form 
\begin{equation}
\label{Defn_eosgammaBar}
\overline{\gamma }=\sum _{j}\Omega _{\phi j}\widehat{\gamma }_{\phi j}+
\gamma _{BG}\left( 1-\sum _{j}\Omega _{\phi j}\right).
\end{equation}

We start by briefly discussing the special case of models with fixed
\( \lambda _{i} \) (i.e.
$ \Gamma _{i} \equiv \frac{V_{i}V_{i,\phi _{i},\phi _{i}}}{V^{2}_{i,\phi _{i}}} =1 $). In that case, the
above system of equations becomes an autonomous system with a set
of {\it true fixed points}, which are obtained by solving  the equations 
\( \frac{dx_{i}}{dN}=\frac{dy_{i}}{dN}=0\; \) 
or \( \frac{d\Omega _{\phi i}}{dN}=\frac{d\widehat{\gamma }_{\phi i}}{dN}=0,\; \forall \, i \).
We have calculated these fixed points for the systems (\ref{xyRhoBg_eom_a})
and (\ref{eom_omegagamma}) and the results are summarized in 
Tables~\ref{TAB_FixPnts_A} and~\ref{TAB_FixPnts_B}.
Included are the existence conditions
as well as the required ranges of the \( \lambda _{i} \) and \( k_{i} \).
These are the generalizations of the fixed points given by~\cite{LiddleTrack},
to the case of models with \( n \) scalar fields and generalized temperature 
dependent potentials (with \( k_{i} \neq 0 \)).
Note that in the single field case, in the \( k_{i}=0 \) limit, points \( A \)
through \( E \) correspond to the fixed points found in~\cite{LiddleTrack},
while fixed points \( F \) and \( G \) are new.

For \( \Omega _{\phi j}\neq 0 \) all the fixed points satisfy
\( \widehat{\gamma }_{\phi j}=\overline{\gamma } \),
which implies \( \left( \overline{\gamma }-\gamma _{BG}\right) \left( 1-\sum _{j}\Omega _{\phi j}\right) =0 \).
As can be seen from Tables~\ref{TAB_FixPnts_A} and~\ref{TAB_FixPnts_B},
the fixed points that are most interesting for cosmological model building
fall into two groups: \( A \) and \( B \). These exist for \( k_{i}\neq 6 \)
and \( \lambda _{i}\neq 0 \) and can be represented by 
\begin{equation}
\label{FixedPnt_A_B_x_y}
\begin{array}{c}
x_{i}=\sqrt{\frac{3}{2}}\frac{\overline{\gamma }-\frac{k_{i}}{3}}{\lambda _{i}}\\
y_{i}=\sqrt{\frac{3}{2}}\frac{\sqrt{\left( 2-\overline{\gamma }\right) \left( \overline{\gamma }-\frac{k_{i}}{3}\right) }}{\lambda _{i}}
\end{array}
\end{equation}
or \begin{equation}
\label{FixPnt_main_Omega}
\begin{array}{cc}
\left( 1-\frac{k_{i}}{6}\right) \left( \widehat{\gamma }_{\phi i}-\frac{k_{i}}{3}\right) =\frac{1}{3}\Omega _{\phi i}\lambda ^{2}_{i},\; \; \;  & \; \; \; \widehat{\gamma }_{\phi i}=\overline{\gamma }
\end{array}
\end{equation}
which are only valid for \( \overline{\gamma }>\frac{k_{i}}{3} \).
Note that the effective equation of state is 
\( \widehat{\gamma }_{\phi i}=\overline{\gamma } \),
which is independent of the value of \( k_{i} \). Interestingly, the actual
equation of state \( \gamma _{\phi i} \) will decrease as \( k_{i} \)
increases so as to maintain \( \widehat{\gamma }_{\phi i}=\overline{\gamma } \):
\begin{equation}
\label{FixPnt_A_B_with_k}
\gamma _{\phi i}=\frac{2x^{2}_{i}}{x^{2}_{i}+y^{2}_{i}}=2\frac{\overline{\gamma }-\frac{k_{i}}{3}}{2-\frac{k_{i}}{3}}
\end{equation}
For all points other
than types \( A \) and \( B \), the equation of state of the field
is fixed to be either \( 2 \) or \( \frac{k_{i}}{3} \), which must
be the same as the average equation of state \( \overline{\gamma } \)
(and the equation of state of the background \( \gamma _{BG} \) if
\( \Omega _{BG}>0 \)). Thus, from the point of view of the other
fields, which see only the expansion rate, one can always absorb the
fields at points other than \( A \) and \( B \) by a redefinition
of the background \( \widetilde{\Omega }_{BG}=\Omega _{BG}+\sum _{j\in \left\{ fixed\: \gamma \right\} }\Omega _{\phi j} \),
and relabeling the \( B \) point by \( A \) if \( \widetilde{\Omega }_{BG}>0 \).
No generality is lost because this redefinition of the background
has no effect on the expansion rate, and it is only through the expansion
rate that one field can affect another. In the following discussion
we shall drop the tilde in \( \widetilde{\Omega }_{BG} \), and by \( \Omega _{BG} \)
we shall mean the total background $\Omega$ plus the contributions from \( \phi _{i} \)
at fixed points other than \( A \) and \( B \). Thus in the light of above discussion,
we shall only concentrate
on the fixed points of type \( A \) and \( B \) and consider
the following two possible scenarios:

\begin{enumerate}
\item \( \widetilde{\Omega }_{BG}>0 \): In this case each field \( \phi_i \) is at the fixed
point of type \( A \). The average equation of state
is the same as the background (\( \overline{\gamma }=\gamma _{BG} \))
and the the fields can 
track the background energy density, making
this point the most relevant for quintessence. The \( \Omega _{\phi i} \)
are in this case given by Eq.~(\ref{FixPnt_main_Omega}) to be 
\begin{equation}
\label{FixPnt_A_Omega}
\Omega _{\phi i}=\frac{3\left( \gamma _{BG}-\frac{k_{i}}{3}\right) \left( 1-\frac{k_{i}}{6}\right) }{\lambda ^{2}_{i}}.
\end{equation}
 For this arrangement to exist as a fixed point one requires \begin{equation}
\label{FixPnt_A_ExstCond}
\Omega ^{(A)}_{\phi }=\sum _{i}3\frac{\left( 1-\frac{k_{i}}{6}\right) \left( \gamma _{BG}-\frac{k_{i}}{3}\right) }{\lambda ^{2}_{i}}<1
\end{equation}
 Note that modestly larger values of the \( k_{i} \) make point \( B \)
less likely and point \( A \) more likely (because it is harder for
the fields, for a given set of \( \lambda _{i} \), to come to dominate
the total energy density). Thus the presence of \( k_{i} > 0 \)
generally makes tracking
(point \( A \)) more robust.
\item \( \widetilde{\Omega }_{BG}=0 \): In this case each field \( \phi_i \) is at the fixed
point of type \( B \) and we have 
\begin{equation}
\label{FixPnt_B_Omega}
\Omega _{\phi i}=3\frac{\left( 1-\frac{k_{i}}{6}\right) \left( \overline{\gamma }-\frac{k_{i}}{3}\right) }{\lambda ^{2}_{i}}
\end{equation}
Now summing over $i$, one finds
\begin{equation}
\label{FixPnt_B_Omega_2}
%\begin{array}{c}
\frac{3\overline{\gamma }}{\mu ^{2}}-6\sigma =1
%\begin{array}{cc}
\end{equation}
where $\sigma$ and $\mu$ are given by
\begin{equation}
\frac{1}{\mu ^{2}}\equiv \sum _{i}\frac{\left( 1-\frac{k_{i}}{6}\right) }{\lambda ^{2}_{i}},\;  ~~~
\sigma \equiv \sum _{i}\frac{k_{i}}{6}\frac{\left( 1-\frac{k_{i}}{6}\right) }{\lambda ^{2}_{i}}.
%\end{array}
%\end{array}
\end{equation}
Note that \( \mu  \) is analogous to \( \lambda  \) 
in the single-field case with
\( k_{i}=0 \). One can make an analogy with parallel resistors
in electrostatics, whereby the smallest \( \lambda _{i}^{2} \) dominates
\( \mu ^{2} \). In the case of $k_i =0$,
the results of previous work concerning  assisted-inflation~\cite{AssistInfl}
are recovered.
Note also that because \( k_{i}\leq 6 \),
\( \mu ^{-2}\geq \sigma \geq 0 \).
However, the ensemble of fixed points does not exist for this entire
range of values: they only exist if \( \overline{\gamma }\in [0,2] \)
and \( \Omega _{\phi j}\in [0,1] \). One has \( \overline{\gamma }=\mu ^{2}\frac{1+6\sigma }{3}\leq 2 \),
which, in turn, fixes all of the \( \Omega _{\phi j} \). This arrangement
does not exist as a fixed point for \( \frac{6}{\mu ^{2}}-6\sigma <1 \)
or \( \frac{k_{i}}{\mu ^{2}}-6\sigma >1 \)
\end{enumerate}

\begin{table}

\caption{\textbf{Points available for \protect\( \Omega _{BG}>0\protect \):}}

\begin{tabular}{|l|c|c|c|c|}
\hline 
\( i \)th point&
 \( x_{i} \)&
 {\footnotesize \( y_{i} \)}&
 {\footnotesize \( \Omega _{\phi i} \)}&
 {\footnotesize \( \widehat{\gamma }_{\phi i}=\overline{\gamma }= \)}\\
\hline
\hline 
\( A \)&
\( \sqrt{\frac{3}{2}}\frac{\gamma _{BG}-\frac{k_{i}}{3}}{\lambda _{i}} \)&
\( \sqrt{\frac{3\left( \gamma _{BG}-\frac{k_{i}}{3}\right) \left( 2-\gamma _{BG}\right) }{2\lambda ^{2}_{i}}} \)&
\( \sqrt{\frac{3\left( \gamma _{BG}-\frac{k_{i}}{3}\right) \left( 2-\gamma _{BG}\right) }{2\lambda ^{2}_{i}}} \)&
\( \gamma _{BG} \)\\
\hline
\( C \)&
 {\footnotesize 0}&
 {\footnotesize 0}&
 {\footnotesize 0}&
 {\footnotesize n/a}\\
\hline
{\footnotesize \( D \),\( E \)}&
 {\footnotesize \( x_{i}=\sqrt{\Omega _{\phi i}}\neq 0 \)}&
 {\footnotesize 0}&
 {\footnotesize Any\( \in \left( 0,1\right]  \)}&
 {\footnotesize 2}\\
\hline
\noindent \centering {\footnotesize \( F \)}&
 {\footnotesize 0}&
 {\footnotesize \( y_{i}=\sqrt{\Omega _{\phi i}}\neq 0 \)}&
 {\footnotesize Any\( \in \left( 0,1\right]  \)}&
 {\footnotesize \( =\frac{k_{i}}{3}=\gamma _{BG} \)}\\
\hline
\( G \)&
 {\footnotesize Any\( \neq 0 \)}&
 {\footnotesize Any\( \neq 0 \)}&
 {\footnotesize Any\( \in \left( 0,1\right]  \)}&
 {\footnotesize 2}\\
\hline
\end{tabular}\label{TAB_FixPnts_A}

{\noindent \centering \begin{tabular}{|l|c|c|c|}
\hline 
\( i \)th point&
{\footnotesize req\( \lambda _{i} \)}&
{\footnotesize req\( k_{i} \)}&
{\footnotesize exist cond.}\\
\hline
\hline 
\( A \)&
{\footnotesize \( \neq 0 \)}&
{\footnotesize \( \neq 6 \)}&
{\footnotesize \( \sum _{j}\Omega _{\phi j}<1 \)}\\
\hline
\( C \)&
 {\footnotesize -}&
 {\footnotesize -}&
 {\footnotesize -}\\
\hline
{\footnotesize \( D \),\( E \)}&
 {\footnotesize -}&
 {\footnotesize -}&
 {\footnotesize \( \overline{\gamma }=2\Rightarrow \begin{array}{cc}
1) & \forall j\; \Omega _{\phi j}\left( \widehat{\gamma }_{\phi j}-2\right) =0\\
2) & \gamma _{BG}=2
\end{array} \)}\\
\hline
\noindent \centering {\footnotesize \( F \)}&
 {\footnotesize 0}&
 {\footnotesize \( =3\gamma _{BG} \)}&
 {\footnotesize \( \sum _{j}\Omega _{\phi j}<1 \) and \( \gamma _{BG}=\overline{\gamma } \)}\\
\hline
\( G \)&
 {\footnotesize 0}&
 {\footnotesize 6}&
 {\footnotesize \( \overline{\gamma }=2\Rightarrow \begin{array}{cc}
1) & \forall j\; \Omega _{\phi j}\left( \widehat{\gamma }_{\phi j}-2\right) =0\\
2) & \gamma _{BG}=2
\end{array} \)}\\
\hline
\end{tabular}\par}
\end{table}

\begin{table}

\caption{\textbf{Points available for \protect\( \Omega _{BG}=0\protect \):}}

\begin{tabular}{|c|c|c|c|c|}
\hline 
\( i \)th point&
 \( x_{i} \)&
 {\footnotesize \( y_{i} \)}&
 {\footnotesize \( \Omega _{\phi i} \)}&
 {\footnotesize \( \widehat{\gamma }_{\phi i}=\overline{\gamma }= \)}\\
\hline
\hline 
\( B \)&
{\footnotesize \( \sqrt{\frac{3}{2}}\frac{\overline{\gamma }-\frac{k_{i}}{3}}{\lambda _{i}} \)}&
 {\footnotesize \( \sqrt{\frac{3\left( \overline{\gamma }-\frac{k_{i}}{3}\right) \left( 2-\overline{\gamma }\right) }{2\lambda ^{2}_{i}}} \)}&
 {\footnotesize \( \frac{3\left( \overline{\gamma }-\frac{k_{i}}{3}\right) \left( 1-\frac{k_{i}}{6}\right) }{\lambda ^{2}_{i}} \)}&
 {\footnotesize \( \frac{1+\sum _{j}\frac{3}{\lambda ^{2}_{j}}\left( 1-\frac{k_{j}}{6}\right) \frac{k_{j}}{3}}{\sum _{j}\frac{3}{\lambda ^{2}_{j}}\left( 1-\frac{k_{j}}{6}\right) } \)} \\
\hline
\( C \)&
 {\footnotesize 0}&
 {\footnotesize 0}&
 {\footnotesize 0}&
 {\footnotesize n/a}\\
\hline
{\footnotesize \( D \),\( E \)}&
 {\footnotesize \( x_{i}=\sqrt{\Omega _{\phi i}}\neq 0 \)}&
 {\footnotesize 0}&
 {\footnotesize Any\( \in \left( 0,1\right]  \)}&
 {\footnotesize 2}\\
\hline
\( F \)&
 {\footnotesize 0}&
 {\footnotesize \( y_{i}=\sqrt{\Omega _{\phi i}}\neq 0 \)}&
 {\footnotesize \( =1-\sum _{j\neq i}\Omega _{\phi j} \)}&
 {\footnotesize \( =\frac{k_{i}}{3} \)}\\
\hline
\( G \)&
 {\footnotesize Any\( \neq 0 \)}&
 {\footnotesize Any\( \neq 0 \)}&
 {\footnotesize Any\( \in \left( 0,1\right]  \)}&
 {\footnotesize 2}\\
\hline
\end{tabular}\label{TAB_FixPnts_B_2}\label{TAB_FixPnts_B}

{\noindent \centering \begin{tabular}{|c|c|c|c|}
\hline 
\( i \)th point&
 {\footnotesize req\( \lambda _{i} \)}&
 {\footnotesize req\( k_{i} \)}&
 {\footnotesize exist cond.}\\
\hline
\hline 
\( B \)&
 {\footnotesize \( \neq 0 \)}&
 {\footnotesize \( \neq 6 \)}&
{\footnotesize \( \begin{array}{c}
\frac{k_{i}}{\mu ^{2}}-6\sigma \leq 1\; \forall i\\
\frac{6}{\mu ^{2}}-6\sigma \geq 1
\end{array} \)}\\
\hline
\( C \)&
 {\footnotesize -}&
 {\footnotesize -}&
 {\footnotesize -}\\
\hline
{\footnotesize \( D \),\( E \)}&
 {\footnotesize -}&
 {\footnotesize -}&
 {\footnotesize \( \overline{\gamma }=2\Rightarrow \forall j\; \Omega _{\phi j}\left( \widehat{\gamma }_{\phi j}-2\right) =0 \)}\\
\hline
\( F \)&
 {\footnotesize 0}&
 {\footnotesize -}&
 {\footnotesize \( \Omega _{\phi i}=1-\sum _{j\neq i}\Omega _{\phi j}>0 \)}\\
\hline
\( G \)&
 {\footnotesize 0}&
 {\footnotesize 6}&
 {\footnotesize \( \overline{\gamma }=2\Rightarrow \forall j\; \Omega _{\phi j}\left( \widehat{\gamma }_{\phi j}-2\right) =0 \)}\\
\hline
\end{tabular}\par}
\end{table}

It turns out that point \( A \) is more interesting for tracking
quintessence models than \( B \), as the latter is generally ruled
out by a number of observational constraints, such as Big bang nucleosynthesis
(BBN) and structure formation~\cite{BBNBound,StrFormBound}.
However, point \( B \) has been studied in scenarios where the domination
of the scalar fields is desirable, such as in the case of assisted inflation~\cite{AssistInfl}.
Point \( A \) exists for \( \overline{\gamma }=\gamma _{BG} \),
while point \( B \) is only stable for \( \overline{\gamma }<\gamma _{BG} \)
(recall we are considering fixed \( \lambda _{i} \) (\( \Gamma _{i}=1 \));
the situation changes significantly when \( \lambda _{i} \) varies,
as will be shown in the next section). 
%These results represent the generalization
%to multiple field models, with non-zero \( k_{i} \), of results given 
%in~\cite{LiddleTrack} for the case of a single field and 
%\( k_{i} =0 \). 
We also note that points \( A \) and \( B \) both require
\( \frac{k_{i}}{3}<\overline{\gamma },\; \forall i \). Of course
if \( \frac{k_{i}}{3}\geq \overline{\gamma } \) for some \( \phi_i \),
that field becomes unstable but will generally have its corresponding \( \Omega _{\phi i}\rightarrow 0 \),
thus making it harmless as a source of instability.

\section{Tracking by shadowing an instantaneous fixed point }
\label{sectioniii}
In this section we study the possibility of tracking in 
presence of variable $\lambda_i$.
% We recall that the qualitative behavior of dynamical systems
%are in general organized by the attractors present in their phase
%spaces, the simplest of which are the fixed (or equilibrium) points. 
Before doing this for the more general system
(\ref{eom_omegagamma}), we
begin by introducing the notion of tracking in the context of the simpler case of
single scalar field models and briefly discuss how tracking might solve the
fine-tuning problem, as was shown in~\cite{RPtrack,LiddleTrack,PJStrack}.
 
\subsection{Tracking in models with constant $\lambda_i$}

Consider the case with one scalar field, \( \phi  \), and with the
usual form of the potential given by \( k=0 \) and  \( \widehat{\gamma }_{\phi i}=\gamma _{\phi } \).
Assuming \( \lambda  \) to be a constant, the fixed points become `true' fixed
points, given by
\begin{equation}
\label{AttrPnts_SingleField}
\begin{array}{ll}
\left( \Omega _{\phi A},\gamma _{\phi A}\right) =\left( \frac{3\gamma _{BG}}{\lambda ^{2}},\gamma _{BG}\right),  & \lambda ^{2}>3\gamma _{BG}\\
\left( \Omega _{\phi B},\gamma _{\phi B}\right) =\left( 1,\frac{\lambda ^{2}}{3}\right),  & \lambda ^{2}<3\gamma _{BG}.
\end{array}
\end{equation}
For large enough \( \lambda  \), the system will be attracted to
point \( A \), and since asymptotically \( \gamma _{\phi }=\gamma _{BG} \),
this offers a plausible solution to the fine-tuning problem. However,
with a constant \( \lambda  \), \( \Omega _{\phi } \) is also  constant,
which implies that one can not have a significant contribution from quintessence to the
present energy density (\( \Omega _{\phi }\sim 0.7 \)) and at the
same time satisfy the nucleosynthesis or structure formation bounds~\cite{BBNBound,StrFormBound}
(\( \Omega _{\phi }\lesssim 0.15 \)). To make the model compatible
with observations, a variable \( \lambda  \) is required which has
decreased from a large value in the early universe to order of a few
today. Making \( \lambda  \) variable, however, means that the fixed
points are no longer true fixed points, and the question becomes what
determines the asymptotic dynamics of the system. An interesting (and useful) feature
of the evolution Eqs.~(\ref{xyRhoBg_eom_a}) and (\ref{eom_omegagamma})
is that they only involve the value of \( \lambda  \) and not its
derivatives. As a result, the rate of change of the phase space vector
(\( \Omega _{\phi },\gamma _{\phi } \)) at a given instant is the
same as its corresponding value in the constant \( \lambda  \) case,
with the same value of \( \lambda  \). In this sense one may talk
about fixed points at a given time, or \textit{instantaneous fixed
points} that would instantaneously act as attractors (or repelers)
for the typical trajectories of the system. One could then imagine
that there exist dynamical settings with slow enough changes in \( \lambda  \)
such that the trajectories shadow or track a moving instantaneous
fixed point (\( A \) or \( B \) in this case). This is essentially
the tracking scenario, which can in principle allow the quintessence
energy density to follow the energy density of the rest of the universe through
most of its history, and also eliminate dependence on initial conditions.
The issue then becomes under what conditions does tracking behavior
occur. It turns out that further conditions are required for the tracking
solutions to exist and be stable. In the case of one scalar field
with \( \Omega _{\phi }\ll 1 \), it has been shown that there exist
class of potentials for which tracking takes place provided the corresponding
\( \Gamma \) satisfies 
\( \Gamma >1 \) and is nearly a constant~\cite{PJStrack}. 

\begin{figure}
{\centering \resizebox*{0.9\textwidth}{!}{\includegraphics{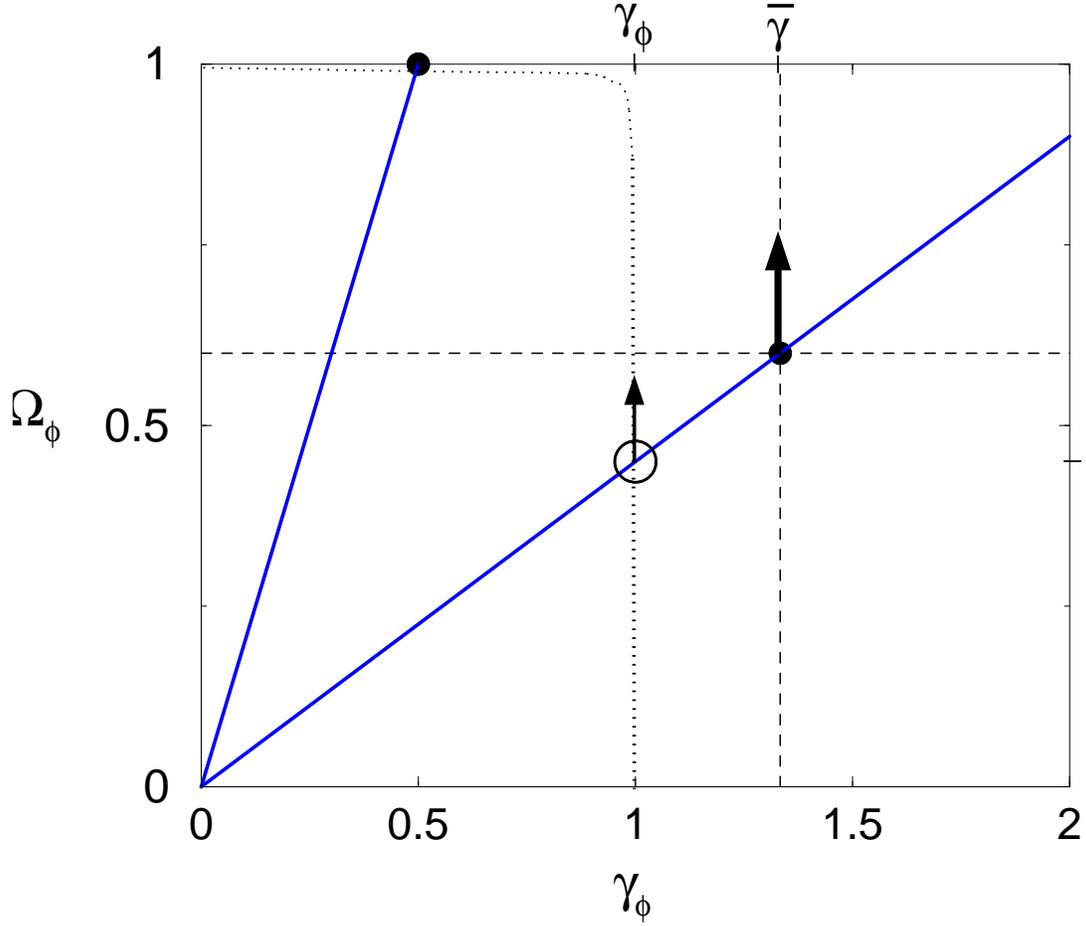}} \par}
\label{FIG1}
\caption{Schematic plot of the 'instantaneous fixed point' (solid dot)
and the the shadowing point (open circle), showing their movement
in the 
\protect\( \left( \gamma _{\phi },\Omega _{\phi }\right) \protect \)
plane. For decreasing \protect\( \lambda _{i}\protect \)
(\protect\( \Gamma _{i}>1\protect \)), the instantaneous fixed point evolves
to larger \protect\( \Omega _{\phi }\protect \). The shadowing point
shadows this point, riding on the \protect\( r=1\protect \)
line. It is at a smaller \protect\( \gamma _{\phi }\protect \) such
that it can stay on the \protect\( r=1\protect \) line as it increases.
Eventually \protect\( \lambda \protect \) will decrease to the point
where $\phi$ dominates (\protect\( \Omega _{\phi }\sim 1\protect \)).}
\end{figure}

\subsection{Tracking in more general settings}

In this section we extend the above analysis of tracking with a single
scalar field and constant \( \lambda \)
to generalized settings with variable \( \lambda _{i} \)'s
as well as 

\begin{enumerate}
\item multiple tracking fields
\item scale-dependent potentials with \( k_{i}\neq 0 \)
\item non-FLRW Hubble laws
(expansion rate different from that of a FLRW universe)
\end{enumerate}

%Though these additions make the resulting models more complicated, it is important
%to consider their effects because in a realistic cosmology motivated
%by high-energy physics they are likely to be present, and because
%their inclusion leads to qualitatively new effects. 

As our interest
is primarily in tracking quintessence models, the emphasis in the
following discussion will be on the shadowing of attractor point \( A \)
in Table~\ref{TAB_FixPnts_A}. As in the case
of models with a single scalar field,
one would expect a stable fixed point to still act as 
an instantaneous type attractor.
When the \( \lambda _{i} \)'s
are allowed to change slowly, the usual
trajectories shadow these points.
This shadowing amounts to the
field being attracted to a surface, defined here by \( r_{i}=1 \).
 From the form of the equation of motion for \( \widehat{\gamma }_{\phi i} \),
it is easy to see 
that \( r_{i}=1 \) is an `attractor surface' (for
\( \widehat{\gamma }_{\phi i}>\frac{k_{i}}{3} \), which we shall
assume), which yields a fixed value of \( \widehat{\gamma }_{\phi i} \).
It is interesting to note that this \( r_{i}=1 \) condition is precisely
equivalent to Eq.~(\ref{FixPnt_main_Omega}), although the latter was
derived as a fixed point for a constant \( \lambda _{i} \) case. We therefore 
focus on tracking that shadows point \( A \) and maintains a constant
equation of state (though in principle other forms of tracking may
be possible) and take 

\begin{equation}
\label{shdpnt_cond_r}
r^{2}_{i}\equiv \frac{\lambda ^{2}_{i}\Omega _{\phi i}}{3\left( \widehat{\gamma }_{\phi i}-\frac{k_{i}}{3}\right) \left( 1-\frac{k_{i}}{6}\right) }=1
\end{equation}
 which ensures \( \widehat{\gamma }_{\phi i}=const \) and in turn
implies \begin{equation}
\label{Tracking_equ_omegaphi}
\Omega _{\phi i}\cong \frac{3}{\lambda ^{2}_{i}}
\left( \widehat{\gamma }_{\phi i}-\frac{k_{i}}{3}\right) \left( 1-\frac{k_{i}}{6}\right).
\end{equation}
For the system of Eqs.~(\ref{eom_omegagamma}) to close we also demand 

\[
\frac{d\ln \Omega _{\phi i}}{dN}\simeq -2\frac{d\ln \lambda _{i}}{dN}\Rightarrow -
2\frac{d\ln \lambda _{i}}{dN}\simeq 3\left( \overline{\gamma }-\widehat{\gamma }_{\phi i}\right). \]

For a FLRW expansion rate, one has\begin{equation}
\label{Tracking_equ_gammaphi}
\frac{d\kappa \phi _{i}}{dN}\simeq \frac{3}{\lambda _{i}}\left( \widehat{\gamma }_{\phi i}-\frac{k_{i}}{3}\right) 
\end{equation}
 which gives \[
-2\frac{d\ln \lambda _{i}}{dN}\simeq -6\left( \widehat{\gamma }_{\phi i}-\frac{k_{i}}{3}\right) \left( 1-\Gamma _{i}\right) \simeq 3\left( \overline{\gamma }-\widehat{\gamma }_{\phi i}\right) \]
which in turn yields an expression for \( \widehat{\gamma }_{\phi i} \) in the form
\begin{equation}
\label{shdpnt_cond_eos}
\widehat{\gamma }_{\phi i}-\frac{k_{i}}{3}\simeq \frac{\overline{\gamma }-\frac{k_{i}}{3}}{2\Gamma _{i}-1}\simeq const.
\end{equation}
Now for \( \widehat{\gamma }_{\phi i} \) to be constant, as is required,
either both \( \overline{\gamma } \) and \( \Gamma _{i} \) must
be constants, or alternatively each needs to be arranged in such 
a way that the ratio
appearing in (\ref{shdpnt_cond_eos}) is constant. But in general there is no
a-priori reason for the latter and we shall therefore not consider this 
possibility further. Since \( \overline{\gamma } \)
is the weighted average of all of the \( \widehat{\gamma }_{\phi i} \),
for tracking to occur for any field, we need either \( \widehat{\gamma }_{\phi j}\simeq const\simeq \overline{\gamma } \)
(which does admit the \( \Omega _{\phi j}=1 \) case) or \( \Omega _{\phi j}\ll 1 \).
 For quintessence, the first case can be consistent with the BBN bound
of \( \Omega _{\phi }\lesssim 0.15 \)~\cite{BBNBound}. However,
if one is considering tracking fields in other scenarios, such as
during inflation, \( \Omega _{\phi }\sim 1 \) can also be acceptable. 

We have given a schematic sketch of the tracking scenario
in Figs. 1 and 2, corresponding to $k=0$ and $k >0$
respectively. In each case we have plotted both
the instantaneous fixed point and the shadowing point
depicted by a solid dot and an open circle respectively.
Note that in each case the shadowing point falls on the $r=1$ line,
but at a shifted $\gamma_\phi$, such that it can remain on
the $r=1$ line, as the system ($\lambda$) evolves.
Figure 2 illustrates the two significant effects
of having $k >0$: the region \( \widehat{\gamma }_{\phi} < k/3  \) is
excluded and the distance between the 
$\gamma_\phi$'s of the instantaneous fixed point and the shadowing point 
is narrower.

In the case of models with a generalized expansion rate
(with \( H_{gen}\neq H_{FLRW} \), as for example
in models which include 
brane corrections or changes in the strength of gravity),
the overall effect
is to change the amount of friction the fields feel. 
More precisely, as was mentioned above, the form of the 
Eq.~(\ref{xyRhoBg_eom_a}) implies that the attractor-type solutions
exist as before, but their properties are determined by a new effective
logarithmic slope $\widetilde{\lambda }_{i}$ defined by Eq.~(\ref{Defn_lambda_eff}).
The \( i-th \) field will then \emph{instantaneously} behave as if
it had this effective logarithmic slope. However,
the evolution of \( \widetilde{\lambda }_{i} \) as a function
of \( N \) is not determined by rescaling \( \lambda _{i}\left( N\right)  \)
by Eq.~(\ref{Defn_lambda_eff}), since the \( \lambda _{i} \) evolution
equation does not transform in this way. As a result, when a correction to the
expansion rate is present, the qualitative nature of the attractor
can be understood from the value (or range of applicable values)
of \( \widetilde{\lambda }_{i} \). The robustness of tracking then
depends on the nature of the modified attractor of the field \( \phi_i \),
which is determined by the value of \( \widetilde{\lambda }_{i} \).
Of course by the time of nucleosynthesis, \( H_{gen}\simeq H_{FLRW} \),
and the attractors will then be unmodified until the present time. 

Whether a given modification to the expansion rate helps or harms
tracking depends on the details of cosmological scenario
in question. For example, a larger expansion rate in the early universe
may cause an attractor to be at a smaller value of the field today
where the slope of the potential is larger, and the field less able
to dominate the energy content of the universe - or it may turn out
that at a smaller field value the slope of the potential is less and
thus the field is more likely to become dominant. Additionally, the
transition from the non-FLRW attractor to the FLRW attractor may be
abrupt, if the field can not shadow the attractor during the transition,
in which case tracking can be disrupted. Thus, the effect of modifications
to the Hubble law on the robustness of tracking depends on the details of
the cosmological model - but given the necessary details, the effect
can be determined by analyzing the attractors resulting from Eq.~(\ref{Defn_lambda_eff})

\begin{figure}
{\centering \resizebox*{0.9\textwidth}{!}{\includegraphics{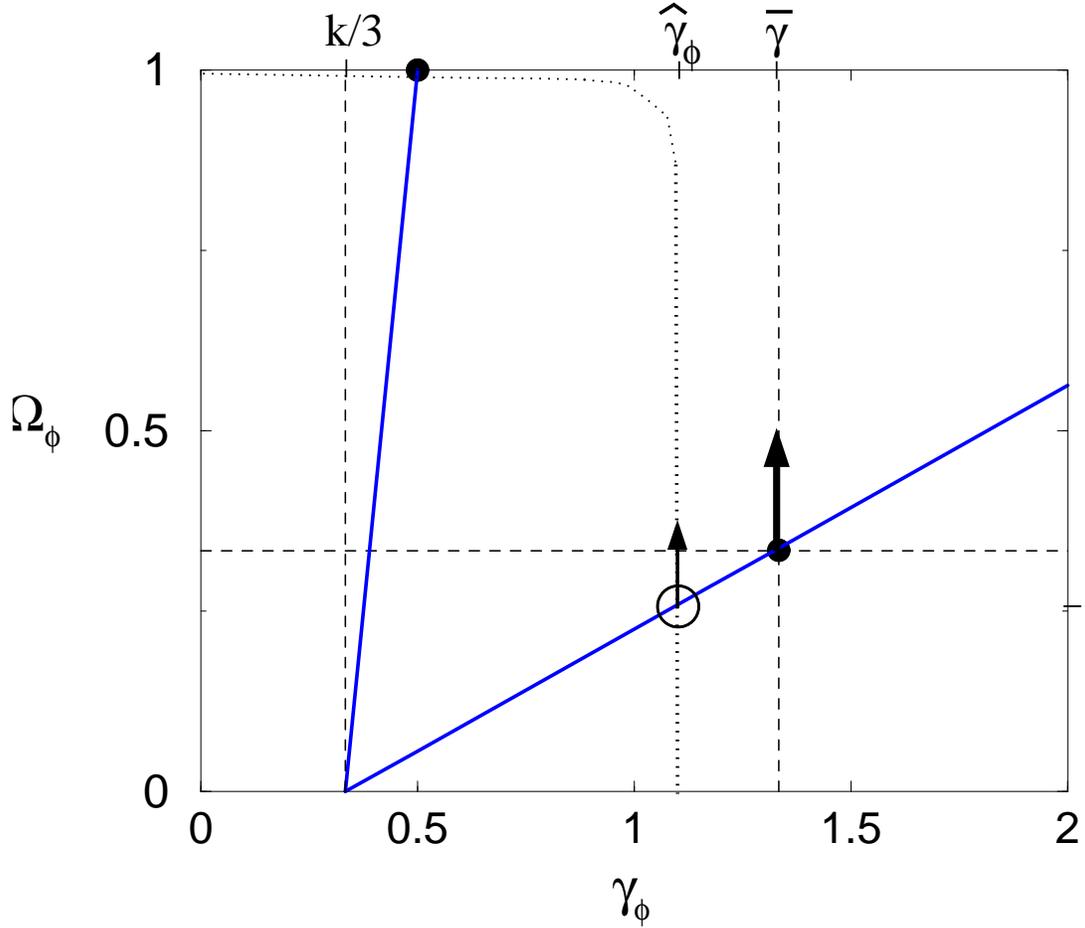}} \par}
\label{FIG2}

\caption{Schematic plot showing the effect of making the field's potential scale factor
dependent (\protect\( V\left( \phi ,N\right) \propto \exp \left( -kN\right) \protect \)).
The effect of \protect\( k>0\protect \) is to set a minimum effective
equation of state (that is, decay rate of \protect\( \rho _{\phi }\protect \))
for the field and to make it larger for the attracting point. Thus the shadowing point
shadows the 'instantaneous fixed point' more closely and comes to dominate more
slowly. However, for \protect\( k\geq 3\overline{\gamma }\protect \)
tracking is impossible and the field dies away (\protect\( \Omega _{\phi }\rightarrow 0\protect \)).
In this way \protect\( k\protect \) can be said to act as a 'throttle'
for quintessence.}
\end{figure}

\subsection{Tracking and independence from initial conditions}

In general, the attractor of field \( \phi _{j} \) is unique and
independent of initial conditions. We will show this by constructing
an equation involving \( \phi _{j} \) and \( \rho _{BG} \) as the
only dynamical variables, which does not include the initial conditions
of any of the fields \( \phi _{j} \). One can then imagine solving
this equation to obtain \( \phi _{j}\left[ \rho _{BG}\right]  \),
or equivalently, \( \phi _{j}\left[ N\right]  \). In general one
would expect this equation to possess a single, monotonically varying
solution - although special cases where this is not the case can undoubtedly
be constructed. The important point is that this solution is independent
of the initial conditions of the fields.

To see this, the first step is to find \( \overline{\gamma } \) as a function
of the field values \( \phi _{j} \). Using its definition and Eqs.~(\ref{Tracking_equ_omegaphi})
and (\ref{Tracking_equ_gammaphi}) for each field, one can find a quadratic
equation for \( \overline{\gamma } \) as a function of \( \lambda _{j} \),
\( \Gamma _{j} \), \( k_{j} \) with \( j=1,n \), in the form
\begin{equation}
\label{Unq_gammaBar}
\overline{\gamma }^{2}\left( \sum _{j}A_{j}\right) -\overline{\gamma }\left( 1+\sum _{j}A_{j}\left( B_{j}+\frac{k_{j}}{3}\right) \right) +\left( \gamma _{BG}+\sum _{j}A_{j}B_{j}\frac{k_{j}}{3}\right) =0
\end{equation}
 where \begin{equation}
\label{Uni_gammaBar_coeff}
\begin{array}{c}
A_{j}\equiv \frac{3\left( 1-\frac{k_{j}}{6}\right) }{\lambda _{j}^{2}\left( 2\Gamma _{j}-1\right) ^{2}}\\
B_{j}\equiv \left( 2\Gamma_{j}-1\right) \left( \gamma_{BG}-\frac{k_{j}}{3}\right) +\frac{k_{j}}{3}.
\end{array}
\end{equation}
As this is a quadratic equation for \( \overline{\gamma }\in \left[ 0,2\right]  \),
there may exist \( 0 \), \( 1 \), or \( 2 \) solutions for \( \overline{\gamma } \).
The existence and stability of such solutions will be dealt with in the next
section. The key point here is that solutions are manifestly independent
of initial conditions. Combining Eqs.~(\ref{Tracking_equ_omegaphi})
and (\ref{Tracking_equ_gammaphi}) yields 
\begin{equation}
\label{Unq_Omega}
\Omega _{\phi i}=\left( \frac{3}{\lambda _{i}^{2}}\right) \frac{\left( 1-\frac{k_{i}}{6}\right) \left( \overline{\gamma }-\frac{k_{i}}{3}\right) }{2\Gamma _{i}-1}=\Omega _{\phi i}\left[ \phi _{k}\right] =\frac{1}{1+\frac{\rho _{BG}}{\rho _{\phi i}}}.
\end{equation}
One can also write an expression for \begin{equation}
\label{Unq_rho_phi}
\rho _{\phi i}=\frac{V}{1-\frac{1}{2}\gamma _{\phi i}}=\frac{V}{1-\frac{\overline{\gamma }-\frac{k_{i}}{3}}{2\left( 2\Gamma _{i}-1\right) \left( 1-\frac{k_{i}}{6}\right) }}=\rho _{\phi i}\left[ \phi _{k}\right] 
\end{equation}
 Thus one arrives at the following equation \begin{equation}
\label{Unq_phi_rhoBG_equ}
\Omega _{\phi i}\left[ \phi _{k}\right] =\frac{1}{1+\frac{\rho _{BG}}{\rho _{\phi i}\left[ \phi _{k}\right] }}
\end{equation}
 This will generally be a very difficult expression to solve. However,
it simplifies greatly when \( \Omega _{\phi i}\ll 1 \), because the
fields are only coupled to each other through the effects each has on the
Hubble expansion rate, and for \( \Omega _{\phi i}\ll 1 \) they effectively decouple,
reducing the complexity of the system. In principal, the solution
(if it exists, is physical, and is stable - issues we shall treat below)
yields \( \phi _{i}\left[ N\right], \; \; i=1,n \), which are independent
of initial conditions. Thus if the solution exists (i.e. is self-consistent)
and is stable, then it is independent of the initial conditions. 

The above argument depends on all of the fields being 'tracking fields'
- that is, Eqs.~(\ref{Tracking_equ_omegaphi}) and~(\ref{Tracking_equ_gammaphi})
being consistent with \( r_{i}=1 \) and thus \( \widehat{\gamma }_{\phi i}=const \).
Of course, not all the fields in a multi-field model need necessarily possess a potential
that is compatible with tracking, in which case those fields would then not track.
In such a case the argument given above for the 
independence from initial conditions
is no longer valid. It is then impossible to make a general prediction
about the existence or initial condition-independence of attractors
for the tracking fields. However, as the fields affect one another
through altering the background expansion rate (or altering the value
and rate of change of \( \overline{\gamma } \)), one expects that
if the \( \Omega _{\phi i} \) of the non-tracking fields stay small,
or if they do not cause \( \overline{\gamma } \) to vary rapidly,
they will not affect the attractor solution of the tracking fields,
and leave intact the above result concerning independence from initial
conditions, for the tracking fields.

\section{Stability of tracking/shadowing solutions}
\label{sectioniv}

We have shown above that the existence of {\it instantaneous} fixed points for
the evolution equations (\ref{eom_omegagamma}) is mathematically
consistent. However, for these points to give rise to tracking behavior
they must also be attractors, that is, the shadowing point, determined
by Eqs.~(\ref{shdpnt_cond_r}) and (\ref{shdpnt_cond_eos}) must be
stable to perturbations. To find conditions for this, we shall obtain perturbation
equations by perturbing the full evolution equations around the position
of the shadowing point. From the nature of the eigenmodes of the perturbation
equations the stability properties of the shadowing point can then be deduced.
However, for models with \( n \) fields, the problem of determining the perturbation
eigenmodes becomes one of finding the eigenvalues for a \( 2n\times 2n \)
matrix, or solving a polynomial of order \( 2n \). Fortunately, an
exact solution is not necessary in order to address several important issues.
These include the stability in the nearly decoupled limit (\( \Omega _{\phi i}\ll 1 \)) and
the determination of when the shadowing point will become unstable. Since the
fields \( \phi _{i} \) only affect each other through the expansion
rate, one would expect that for small \( \Omega _{\phi i}\ll 1 \)
the system will behave as \( n \) decoupled systems, which turns out to be the case.
The stability conditions for the decoupled fields correspond to the generalization
of those
found in~\cite{LiddleTrack}, with the added explicit scale
dependence of the potential or corrections to the Hubble law taken
into account. For the coupled case the results are also fairly intuitive:
from the the shadowing conditions, one can see that \( \widehat{\gamma }_{\phi i} \)
should be nearly constant, and equal to a nearly constant multiple of the overall
average equation of state \( \overline{\gamma } \). Thus we have
a key stability condition, namely that \( \overline{\gamma } \) must be nearly
constant. Now when \( \Omega _{BG} \) dominates this condition is satisfied.
However, if \( \Omega _{\phi i}\left( \Gamma _{i}-1\right)  \)
become of order unity for any field $\phi_i$, even if each \( \widehat{\gamma }_{\phi i} \)
is constant, the proportion of the total energy density in the field
increases rapidly, resulting in a rapid
decease in \( \overline{\gamma } \).
Similarly, as \( \sum _{j}\Omega _{\phi j} \)
approaches unity, the condition \( \overline{\gamma }=\gamma _{BG}\simeq const \)
(and thus \( \widehat{\gamma }_{\phi i}\simeq const \)) can no longer be
maintained. As one would expect, this 
manifests itself as a growing mode of the perturbations, 
which amounts to an instability. This disrupts tracking
as \( \sum _{j}\Omega _{\phi j}\rightarrow 1 \).
The important question is whether perturbations have growing modes if \( \sum _{j}\Omega _{\phi j} < 1 \)?
In the next section we find that the answer to this question
is negative, so long as \( \overline{\gamma } \) and $\Gamma_i$ remain nearly
constant.

\subsection{The perturbation equations}

Assume for the moment that all fields are trackers - that is expressions (\ref{shdpnt_cond_r})
and (\ref{shdpnt_cond_eos}) hold for each field individually. Now to study the perturbations,
take
\begin{equation}
\label{Stab_PertDefns}
\begin{array}{ll}
r_{i}-1=\epsilon _{i},\; \; \; \; \;  & \; \; \; \; \; 
\left( 2\Gamma _{i}-1\right) \left( \widehat{\gamma }_{\phi i}-\frac{k_{i}}{3}\right) =-\delta _{i}+\left( \overline{\gamma }-\frac{k_{i}}{3}\right) 
\end{array}
\end{equation}
with the perturbations \( \epsilon _{i},\delta _{i}\ll 1 \). 
The resulting evolution equations for these perturbations 
become
\begin{equation}
\label{Stab_Pert_Evol_eqs}
\begin{array}{c}
\frac{d\epsilon _{i}}{dN}=\frac{3}{2}\delta _{i}-3\epsilon _{i}\left( 2+\overline{\gamma }\frac{2\Gamma _{i}-3}{2\Gamma _{i}-1}-4\frac{k_{i}}{3}\frac{\Gamma _{i}-1}{2\Gamma _{i}-1}\right) +\Delta \overline{\gamma }\\
\\
\frac{d\delta _{i}}{dN}=-\epsilon _{i}3\left( \overline{\gamma }-\frac{k_{i}}{3}\right) \left( \left( 2-\frac{k_{i}}{3}\right) -\frac{\overline{\gamma }-\frac{k_{i}}{3}}{2\Gamma _{i}-1}\right) \\
+\frac{d\overline{\gamma }}{dN}-\frac{2\left( 1-\delta _{i}\right) }{2\Gamma _{i}-1}\frac{d\Gamma _{i}}{dN},
\end{array}
\end{equation}
where the perturbation in \( \overline{\gamma } \) due to the perturbation
of each field is given by
\begin{equation}
\label{Stab_Deltagamma}
\begin{array}{c}
\Delta \overline{\gamma }\equiv \frac{3}{2}\sum _{j}\Omega _{\phi j}\left[ \delta _{j}\left( \frac{\gamma _{BG}-\frac{k_{j}}{3}}{\overline{\gamma }-\frac{k_{j}}{3}}-\frac{2}{2\Gamma _{j}-1}\right) \right. \\
\left. -\epsilon _{j}2\left( \left( \gamma _{BG}-\frac{k_{j}}{3}\right) -\frac{\overline{\gamma }-\frac{k_{j}}{3}}{2\Gamma _{j}-1}\right) \right].
\end{array}
\end{equation}
This is the source of the coupling of the perturbations. Furthermore,
the rate of change of \( \overline{\gamma } \) enters into the perturbation
equation, which is given by
\begin{equation}
\label{Stab_gammabar_variation}
\frac{d\overline{\gamma }}{dN}=\sum _{j}\Omega _{\phi j}\left[ \left( \Gamma _{j}-1\right) 6\left( \widehat{\gamma }_{\phi j}-\frac{k_{j}}{3}\right) \left( \widehat{\gamma }_{\phi j}-\gamma _{BG}\right) +\frac{d}{dN}\left( \widehat{\gamma }_{\phi j}-\gamma _{BG}\right) \right] +\frac{d\gamma _{BG}}{dN}.
\end{equation}
Note that we are not presently taking \( \gamma _{BG} \) to be a constant
- the reason for which will become clear when we consider the effect on stability
by non-tracking fields by absorbing them into a redefinition of the
background. The condition for the closing of the unperturbed system of equations
is that the terms in the stability equation above that are not
proportional to \( \epsilon _{j} \) or \( \delta _{j} \) must vanish.
Thus the following terms:
\begin{equation}
\label{Stab_NonClosure_terms}
%\begin{array}{c}
\sum _{j}\Omega _{\phi j}\left( \Gamma _{j}-1\right), ~~
\frac{d\ln \left( 2\Gamma _{i}-1\right) }{d\kappa \phi _{i}}, ~~
\frac{d\overline{\gamma }}{dN}
%\end{array}
\end{equation}
 must be negligible compared to \( \epsilon _{j} \)
or \( \delta _{j} \).
 For the \( i-th \) field \( \phi _{i} \) to have a stable \( r_{i}=1 \)
tracking attractor, one requires that \( \frac{d\ln \left( 2\Gamma _{i}-1\right) }{d\kappa \phi _{i}}\ll 1\; \)
for all $i$ as well as
\( \sum _{j}\Omega _{\phi j}\left( \Gamma _{j}-1\right) \ll 1 \)
and \( \frac{d\gamma _{BG}}{dN}\ll 1 \). That is, if the \( \epsilon _{i} \)
and \( \delta _{i} \) equations without these terms are stable (have
only decaying modes, with negative real parts of eigenvalues) then
the \( \epsilon _{i} \) and \( \delta _{i} \) will decay until they
are of the order of these 'residual' quantities. Alternatively,
one can say that the \( r_{i}=1 \) shadowing point is shifted
by a small amount of the order these terms. Thus we proceed by assuming
the terms that are not proportional to \( \epsilon _{j} \) or \( \delta _{j} \)
are much smaller than \( \epsilon _{j} \) and \( \delta _{j} \),
and analyze the resulting stability equation with these terms removed. This results
in the following equations for \( \epsilon_{i} \) and \( \delta_{i} \)
\begin{equation}
\label{pert_evol_eq1}
\begin{array}{c}
\frac{d\epsilon _{i}}{dN}=\delta _{i}r+\epsilon _{i}a_{i}+\sum _{j}\epsilon _{j}b_{j}+\delta _{j}d_{j}\\
\frac{d\delta _{i}}{dN}=\epsilon _{i}c_{i}
\end{array}
\end{equation}
 with \begin{equation}
\label{stab_perteq_coeff}
\begin{array}{c}
a_{k}\equiv -\frac{3}{2}\left( 2+\overline{\gamma }\frac{2\Gamma _{k}-3}{2\Gamma _{k}-1}-4\frac{k_{k}}{3}\frac{\Gamma _{k}-1}{2\Gamma _{k}-1}\right) \\
b_{k}\equiv -3\Omega _{\phi k}\left( \gamma _{BG}-\frac{k_{k}}{3}-\frac{\overline{\gamma }-\frac{k_{k}}{3}}{2\Gamma _{k}-1}\right) \\
c_{k}\equiv -3\left( \overline{\gamma }-\frac{k_{k}}{3}\right) \left( 2-\frac{k_{k}}{3}-\frac{\overline{\gamma }-\frac{k_{k}}{3}}{2\Gamma _{k}-1}\right) \\
d_{k}\equiv -\frac{3}{2}\Omega _{\phi k}\left( \frac{2}{2\Gamma _{k}-1}-\frac{\gamma _{BG}-\frac{k_{k}}{3}}{\overline{\gamma }-\frac{k_{k}}{3}}\right) 
\end{array}
\end{equation}
and
\[
\Delta \overline{\gamma }=\sum _{j}\epsilon _{j}b_{j}+\delta _{j}d_{j}\]

\noindent In general, finding the eigenmodes of the above perturbation equations
is equivalent to diagonalizing a \( 2n\times 2n \) matrix, or finding
the roots of a polynomial of order \( 2n \) given by
 \begin{equation}
\label{stab_eigenval_poly}
1=\sum _{k}\frac{\beta b_{k}+d_{k}c_{k}}{\beta \left( \beta -a_{k}\right) -\frac{3}{2}c_{k}}.
\end{equation}
Finding explicit expressions for the eigenmodes in terms
of the coefficients (\ref{stab_perteq_coeff}) is neither feasible
nor useful. Even for the simple 2-field case, one is faced with extremely
messy expressions for the roots of a quartic equation. One, however, does not
need to solve for all the \( \beta  \)s explicitly. It is sufficient to 
impose the condition that the real parts of all \emph{relevant} \( \beta  \)s
be negative. A mode is only relevant if its growth leads to the failure
of tracking of the whole system. One would expect that some
fields may not track, and at the same time not interfere with the tracking
of other fields
if they effectively decouple, having \( \Omega _{\phi i}\ll 1 \).
In that case, these are irrelevant, or
harmless instability modes. As a result, \( Re\left[ \beta \right] >0 \)
may not necessarily signal an instability in the entire tracking system,
as the fields only interact by altering the value of \( \overline{\gamma } \),
and with \( \Omega _{\phi i}\ll 1 \) the failure of \( \phi _{i} \)
to track would not effect tracking by the other fields. 

A possible way of studying the stability of the tracking is to recast
the perturbation equations (\ref{pert_evol_eq1}) into a system of equations 
analogous to \( n \) coupled, damped harmonic oscillators (\( \delta _{i} \))
in the form
\begin{equation}
\label{PertEOM_oscmain}
\frac{1}{\left( -c_{i}\right) }\frac{d^{2}\delta _{i}}{dN^{2}}+\frac{\left( -a_{i}\right) }{\left( -c_{i}\right) }\frac{d\delta _{i}}{dN}+\frac{3}{2}\delta _{i}=F\equiv -\sum _{j}\epsilon _{j}b_{j}+\delta _{j}d_{j}.
\end{equation}
If one identifies \( N \) with time, then the oscillator energy is
a good measure of the deviation of the tracking fields from their
shadowing condition (background) values. Thus there are \( n \) oscillators
\( \delta _{i} \), each of which can be excited in \( 2n \) possible
modes. Note that due to its form, equation (\ref{PertEOM_oscmain})
is \emph{not} in general derivable from a conservative Lagrangian.
In particular, note that the coefficients of the terms in \( F \)
proportional to \( \delta _{j} \) do not in general form a symmetric
matrix (and those proportional to \( \epsilon _{j} \) an antisymmetric
matrix), as would be necessary for \( F \) to be derivable from an
interaction potential of the form
\[
F\neq \frac{d}{dN}\frac{\partial V}{\partial \dot{\delta }_{i}}-\frac{\partial V}{\partial \delta _{i}},\; \; \; \; V\equiv \sum _{j,k}C_{(jk)}\delta _{j}\delta _{k}+B_{\left[ jk\right] }\delta _{j}\dot{\delta }_{k}\]
Thus the equations of motion do not conserve 'energy' - that is, the
perturbation amplitudes may decay or grow. However, by examining
the implications of equation (\ref{PertEOM_oscmain}), one can determine
what will happen - and thereby determine the stability of the system.

To do this, we shall employ the ansatz \( \delta _{i}=\sum _{l}A_{il}e^{\beta _{l}N} \)
and then solve mode-by-mode to obtain
\begin{equation}
\label{PertEOM_oscmode}
\frac{A_{il}}{\left( -c_{i}\right) }\left[ \beta ^{2}_{l}-a_{i}\beta _{l}-\frac{3}{2}c_{i}\right] =
F_{l}\equiv \sum _{j}\frac{A_{jl}}{\left( -c_{j}\right) }\left[ d_{j}c_{j}+b_{j}\beta _{l}\right].
\end{equation}
Because the right-hand side is independent of \( i \), we can immediately
determine the relative amplitudes for mode~\( l \) of oscillators~\( i \)
and~\( k \):
\begin{equation}
\label{PertEOM_oscRelAmp}
\frac{A_{il}}{\left( -c_{i}\right) }\left[ \beta ^{2}_{l}-a_{i}\beta _{l}-\frac{3}{2}c_{i}\right] =F_{l}=\frac{A_{kl}}{\left( -c_{k}\right) }\left[ \beta ^{2}_{l}-a_{k}\beta _{l}-\frac{3}{2}c_{k}\right] \; \; \; \; \; \; i\neq k
\end{equation}

We begin by examining the decoupled system (\( F\rightarrow 0 \)).
 For the case of models with \( n \) scalar fields, there are \( 2n \) eigenvalues \( \beta ^{0\pm }_{l} \),
given by
\begin{equation}
\label{Eigenval_0_defn}
\beta ^{0\pm }_{l}=\frac{1}{2}\left( a_{l}\pm \sqrt{a^{2}_{l}+6c_{l}}\right).
\end{equation}
 From Eq.~(\ref{PertEOM_oscRelAmp}) it is clear that for the \( \delta _{i} \)
oscillator only the modes with frequencies \( \beta ^{0\pm }_{i} \)
are present. The zeroth order stability condition is the requirement
that \( Re\left[ \beta ^{0\pm }_{i}\right] <0 \). Using (\ref{Eigenval_0_defn}),
this can be seen to be satisfied if and only if both \( a_{i}<0 \)
and \( c_{i}<0 \). Solving these inequalities yields the following
requirements for stability of uncoupled tracking 
\begin{equation}
\label{Stab_cond_decoupled}
\begin{array}{ccc}
 & \frac{k_{i}}{3}<\overline{\gamma },\; \; \;  & 2\Gamma _{i}-1>\frac{2\left( \overline{\gamma }-\frac{k_{i}}{3}\right) }{2+\overline{\gamma }-2\frac{k_{i}}{3}}>0.
\end{array}
\end{equation}
Of course, for the zeroth order stability conditions to be relevant, we need
\( \sum _{j}\Omega _{\phi j}\left( \Gamma _{j}-1\right) \ll 1 \)
and \( \sum _{j}\Omega _{\phi j}\ll 1 \). Note that for \( k_{i}=0 \),
the stability condition becomes \( 2\Gamma _{i}-1>\frac{2\overline{\gamma }}{2+\overline{\gamma }} \)
or \( \gamma _{\phi _{i}}<\frac{2+\overline{\gamma }}{2} \), which
for \( \overline{\gamma }\approx \gamma _{BG} \) agrees with that
previously obtained in~\cite{PJStrack} for the case of a single tracking field.
This is not surprising, as in the decoupled limit of \( \sum _{j}\Omega _{\phi j}\ll 1 \)
each field can be treated individually. This lower limit on \( 2\Gamma _{i}-1 \)
varies between \( 1 \) and \( \frac{1}{2} \), where \( \Gamma _{i}=1 \)
is allowed unless \( \overline{\gamma }=2 \) and \( \Gamma _{i}<1 \)
corresponds to a diminishing tracking field (\( \frac{d\Omega _{\phi i}}{dN}<0 \)).
 For \( \phi _{i} \) to be a {}``useful quintessence'' field, that
is to have \( \frac{d\Omega _{\phi i}}{dN}>0 \) (\( \Gamma _{i}>1 \)),
the condition (\ref{Stab_cond_decoupled}) is automatically satisfied.
 For \( k_{i}>0 \) and \( \Gamma _{i}>1 \) tracking is still stable,
and the lower limit on \( 2\Gamma _{i}-1 \) becomes smaller, taking 
a value closer to \( \frac{1}{2} \). One can understand this intuitively
as follows: \( k_{i}>0 \) is another channel for the decay of (potential)
energy in the field. Thus as less energy goes into kinetic energy, \( \lambda _{i} \)
changes less rapidly, and the system is more able to maintain the
\( r_{i}=1 \) condition.

We now consider the effect of the coupling between the modes given
by \( F\neq 0 \) in Eq.~(\ref{PertEOM_oscmode}). The coupling causes
the mixing of the modes, but diagonalizing the coupled system, as we have
seen, is not practical. This is not necessary, however, as by examining
the un-diagonalized system we can draw the important conclusion that
the shadowing condition is stable until \( \sum _{k}\Omega _{\phi k}\rightarrow 1 \).
 From Eq.~(\ref{PertEOM_oscRelAmp}) it is clear that for each oscillator
\( \delta _{i} \), the modes with frequencies \( \beta ^{\pm }_{i} \)
are dominant for \( F\ll 1 \), but there are now also \( 2\left( n-1\right)  \)
other modes present, with the frequencies of the other oscillators.
The amplitudes of these other modes are suppressed by a factor of
order \( \sum _{k}\Omega _{\phi k} \) (unless \( \beta ^{0\pm }_{i}\simeq \beta ^{0\pm }_{j} \),
in which case \( A_{ii}\sim A_{jj} \)). 
One can see from Eq.~(\ref{PertEOM_oscmode}) that 
the diagonal mode \( A_{ii} \) can be treated as an oscillator, but with a
shifted \( mass^{2} \). We identify the zeroth order mode's
\( mass^{2} \) with the \( \frac{3}{2}\left( -c_{i}\right) >0 \)
term in Eq.~(\ref{PertEOM_oscmode}). The effect of the interactions
is then to provide a \( mass^{2} \) shift of
\begin{equation}
\label{Pert_DeltaMass2}
\Delta m_{ii}^{2}=\sum _{j}\frac{\beta ^{2}_{i}-a_{i}\beta _{i}-\frac{3}{2}c_{i}}{\beta ^{2}_{i}-a_{j}\beta _{i}-\frac{3}{2}c_{j}}\left( -d_{j}c_{j}-b_{j}\beta _{i}\right).
\end{equation}
The condition for stability then becomes the requirement that the
shifted mass not cause the mode to become growing, which implies 
\( Re\left[ \beta _{i}\right] <0 \). If \( \Delta m_{ii}^{2} \)
was real, then this would be the condition that the shifted \( mass^{2} \)
must be positive
\begin{equation}
\label{Pert_shiftedmass2_cond}
m^{2}_{tot_{ii}}\equiv \frac{3}{2}\left( -c_{i}\right) +\Delta m_{ii}^{2}>0.
\end{equation}
However, in general \( \Delta m_{ii}^{2} \) is complex since
\( \beta _{j} \) are complex. The stability condition then becomes
\begin{equation}
\label{Pert_shiftedmass2_cond_cplx}
Re\left[ m^{2}_{tot_{ii}}\right] -\frac{1}{2}Im\left[ m^{2}_{tot_{ii}}\right]^{2}>0.
\end{equation}
This consideration does not significantly alter the story - that there
is a stability radius of order unity around \( m^{2}_{ii}=\frac{3}{2}\left( -c_{i}\right)  \)
in the complex plane. Thus in this way one can see that
the system is guaranteed to be stable for any parameter values that ensure 
the inequality \( \left| \Delta m_{ii}^{2}\right| <m_{ii}^{2} \)
to be roughly satisfied. 

Thus far we have been considering the diagonal elements of a mass
matrix. We now consider the off-diagonal elements which correspond to
the off-diagonal modes, whose amplitudes are controlled by Eq.~(\ref{PertEOM_oscRelAmp}).
Until \( \sum _{k}\Omega _{\phi k}\sim 1 \), these off-diagonal mode
amplitudes can not become of the same order of the diagonal mode amplitudes, and
therefore Eq.~(\ref{Pert_shiftedmass2_cond_cplx}) remains satisfied. We have seen
that for \( \sum _{j}\Omega _{\phi j}<1 \) the diagonal mode amplitudes
do not grow, and the off-diagonal mode amplitudes are suppressed relative
to these by a factor of order \( \sum _{k}\Omega _{\phi k} \). The
implication is that the shadowing conditions (\ref{shdpnt_cond_r}) and
(\ref{shdpnt_cond_eos}) will be stable so long as \( \sum _{j}\Omega _{\phi j} <1\).

The question remains as to what happens if 
some fields in the model are not trackers, and therefore do
not satisfy the conditions (\ref{shdpnt_cond_r}) and (\ref{shdpnt_cond_eos})?
These fields must then be excluded from the set of 
perturbations \( \left\{ \delta _{i},\epsilon _{i}\right\}  \)
and therefore do not contribute to \( \Delta \overline{\gamma } \).
Because they do not obey the tracking conditions, little can be said
about such fields in general. They do, however, contribute to \( \gamma _{BG} \).
If their \( \Omega  \) is large enough such that they can cause
an appreciable variation of \( \overline{\gamma } \), 
they will cause a deviation
from the shadowing conditions as noted above for the residual terms
(\ref{Stab_NonClosure_terms}).
%-------------------------------------
\subsection{Summary of stability conditions}
%-------------------------------------
We have shown that tracking with the \( r_{i}=1 \) condition is possible
for nearly constant \( \overline{\gamma } \). The constancy of \( \overline{\gamma } \)
is the key to closing the zeroth order tracking equations, and therefore 
it is not surprising that a constant \( \overline{\gamma } \) is the key
to the stability of the shadowing points, and that many of the stability
conditions can ultimately be traced back to it. We note that
even though in principle other forms
of tracking may be possible
under other conditions, the type of tracking considered
here is that which has been commonly
considered in the literature.

%To summarize, we have shown that \( r_{i}=1 \) tracking for all fields requires
Note the near constancy of \( \overline{\gamma } \) in turn requires
nearly constant \( \Gamma _{i} \) together with \( \sum _{i}\Omega _{\phi i}\left( \Gamma _{i}-1\right) \ll 1 \)
and \( \sum _{i}\Omega _{\phi i}<1 \). We have seen that the perturbations
\( \epsilon _{i} \) and \( \delta _{i} \)
will decay until they are of order \( \frac{d\ln \left( 2\Gamma _{i}-1\right) }{d\kappa \phi _{i}} \),
\( \sum _{j}\Omega _{\phi j}\left( \Gamma _{j}-1\right)  \) or \( \frac{d\overline{\gamma }}{dN} \);
resulting in the \( r_{i}=1 \) shadowing point to be shifted by small amount
of the order of these terms. The system is more stable in the
small \( \Omega _{\phi i} \) limit, since as \( \sum _{j}\Omega _{\phi j}\left( \Gamma _{j}-1\right)  \)
or \( \sum _{j}\Omega _{\phi j} \) approach unity, this in general
causes \( \overline{\gamma } \) to vary rapidly, thus making the \( r_{i}=1 \)
condition to fail and the tracking to cease. The weakly
coupled case is analogous to \( n \) individual tracking fields,
and thus it is not surprising that the constraints in the limit of
\( k_{i}\rightarrow 0 \) reproduce those given in~\cite{PJStrack},
for the single field case. Furthermore,
in this limit \( \overline{\gamma }\approx \gamma _{BG} \), and the stability
of the shadowing condition for the field \( \phi _{i} \), given by  (\ref{shdpnt_cond_r}) and (\ref{shdpnt_cond_eos})
require 
\begin{equation}
\label{Cond_0Order_Stable}
\begin{array}{ccc}
\frac{k_{i}}{3}<\overline{\gamma }, & \; \; \; \; \; 2\Gamma _{i}-1>\frac{2\left( \overline{\gamma }-\frac{k_{i}}{3}\right) }{2+\overline{\gamma }-2\frac{k_{i}}{3}}>0.\; \; \; \; 
\end{array}
\end{equation}
If the stability conditions are not met for the field \( \phi_i \), then clearly it
will not track. However, the instability is irrelevant if \( \Gamma _{i}<1 \).
This is because for \( \Gamma _{i}<1 \), the \( \Omega _{\phi i} \)
decreases with expansion (as \( \widehat{\gamma }_{\phi i}>\overline{\gamma } \))
and the field \( \phi_i \) will effectively become irrelevant as \( \Omega _{\phi i}\rightarrow 0 \).
Thus in this case the tracking of the other fields will not be effected by the failure of the 
field \( \phi_i \) to track.
On the other hand, if \( \Gamma _{i}>1 \), or if \( \Gamma _{i}=1 \) and \( \Omega _{\phi i} \)
is of order 1, then the instability in the field \( \phi_i \)  will ultimately ruin tracking
for all fields, as it will cause \( \overline{\gamma } \) to vary
rapidly, and \( r_{i}=1 \) can no longer be maintained in general for any
of the other fields. Finally, we have seen that the presence of non-tracking
fields will harm tracking if and only if they cause \( \overline{\gamma } \)
to vary rapidly.

We note that our stability analysis was limited to the violation of the shadowing
conditions given by (\ref{shdpnt_cond_r}) and (\ref{shdpnt_cond_eos}).
A concern that we have not addressed is the possibility that the homogeneous distribution
of the scalar field may be unstable to formation of spatial inhomogeneities
- the so called 'Q-balls'~\cite{Qball_Paper}. This is an important issue that
we hope to return to in future work.
%----------------------------------
\section{Conclusions}
\label{sectionv}
%----------------------------------
We have argued that for tracking quintessence to truly solve the fine-tuning problem,
the tracking phenomena must be robust in the more complicated and
complete cosmological scenarios that derive from high energy physics.
Such scenarios typically include multiple scalar fields (some of which
may not have potentials suitable for tracking), a Hubble law at high energies different
from that of the FLRW model, and scalar field potentials
that depend explicitly on the scale factor. Tracking would be of limited
use as a solution to the fine-tuning problem if it were easily ruined
by such effects.

In models with $n$ scalar fields, we have found that \( r_{i}=1 \) tracking requires a nearly constant \( \overline{\gamma } \).
If any one of the tracking fields develops a large enough \( \Omega _{\phi i}\left( \Gamma _{i}-1\right)  \),
or if \( \sum _{j}\Omega _{\phi j}\rightarrow 1 \) then in general
\( \overline{\gamma } \) will vary rapidly, causing the tracking to fail
for all fields. Also, if any of the non-tracking fields become a significant
portion of the total energy density then they can potentially cause
\( \overline{\gamma } \) to vary rapidly. In general, therefore, \( r_{i}=1 \)
tracking requires \( \Gamma _{i} \) to be nearly constant for the field
\( \phi_i \) to track, as well as \( \Omega _{\phi i}\left( \Gamma _{i}-1\right) \ll 1 \)
for all fields, and for \( \sum _{j}\Omega _{\phi j} \) not to approach
\( 1 \).

In models with a modified expansion rate (such as those including
the brane corrections or changes in the strength of gravity), we have found
the attractor-type solutions can still exist, but their instantaneous nature
is determined by the effective logarithmic slope \( \widetilde{\lambda }_{i} \).
When such a correction is present, the qualitative nature of the attractor
can be understood from the value of \( \widetilde{\lambda }_{i} \).
However, the answer to the question of whether a given modification to the expansion
rate helps or harms the robustness of tracking depends on the details of
cosmological scenario in question - but given the necessary details,
the effect can be determined by analyzing the attractors resulting
from Eq.~(\ref{Defn_lambda_eff}).

In models with temperature dependent potentials (\( k_{i}>0 \)), we
have found that this dependence can make tracking more robust.
Roughly speaking, \( k_{i}>0 \) 'throttles'
quintessence - slowing the rate of increase of \( \Omega _{\phi i} \),
thus making fields with larger \( \Gamma _{i} \) less harmful to
both tracking stability, and to cosmological scenarios. 
On the other hand, very large values of \( k_{i} \)
spoil tracking for field \( \phi _{i} \). If \( \Gamma _{i}>1 \),
a value of \( 0<k_{i}<3\overline{\gamma } \) keeps (\( \overline{\gamma }-\widehat{\gamma }_{\phi i} \))
small, and slows the increase of \( \Omega _{\phi i} \), making tracking
of the entire system more stable. For \( k_{i}\geq 3\overline{\gamma } \)
the field \( \phi _{i} \) will fail to track, though it becomes an
irrelevant field and tracking of the other fields is not harmed.

In conclusion, we have found that the conditions for robustness of tracking 
becomes more complex in the more general settings that derive from 
high energy physics. Tracking can become more fragile 
with respect to some such added complexities to the model, as there are 
more constraints to be satisfied. This can for example be seen
from the above analysis of stability in presence of multiple
scalar fields. Other additions, such as temperature dependent potentials,
on the other hand 
may make tracking more robust.
Interestingly though, most of the cases where tracking is disrupted 
are those in which the cosmological model is itself non-viable due to other
constraints.
 For example, although tracking is less
robust for larger \( \sum _{j}\Omega _{\phi j} \), this would generally
make the model non-viable due to constraints 
such as nucleosynthesis and structure formation. 
Thus  tracking seems to fair well
in these general settings, once we confine ourselves to
viable cosmological models.

\end{document}